\documentclass[conference]{IEEEtran}
\IEEEoverridecommandlockouts
\usepackage{cite}
\usepackage{amsmath,amssymb,amsfonts}
\usepackage{algorithmic}
\usepackage{graphicx}
\usepackage{textcomp}
\usepackage{xcolor}
\usepackage{url}
\usepackage{booktabs}
\usepackage{multirow}
\def\BibTeX{{\rm B\kern-.05em{\sc i\kern-.025em b}\kern-.08em
    T\kern-.1667em\lower.7ex\hbox{E}\kern-.125emX}}

\makeatletter
\def\ps@IEEEtitlepagestyle{%
  \def\@oddhead{\hbox{}\hfill
    \parbox[b]{\textwidth}{\centering\footnotesize\itshape
    This paper is submitted to the IEEE International Conference on Trust, Security and Privacy in Computing and Communications (TrustCom) 2025 and may be revised following the review process.\par}\hfill\hbox{}}%
  \def\@oddfoot{}%
  \def\@evenhead{\@oddhead}%
  \def\@evenfoot{}%
}
\makeatother

\begin{document}

\title{AutoPentester: An LLM Agent-based Framework for Automated Pentesting}

\author{
\IEEEauthorblockN{
    Yasod Ginige\IEEEauthorrefmark{1},
    Akila Niroshan\IEEEauthorrefmark{2},
    Sajal Jain\IEEEauthorrefmark{3},
    Suranga Seneviratne\IEEEauthorrefmark{1}
}
\IEEEauthorblockA{
    \IEEEauthorrefmark{1}University of Sydney, Australia \\
    \IEEEauthorrefmark{2}University of New South Wales, Australia \\
    \IEEEauthorrefmark{3}Catharsis Pvt. Ltd., Australia \\
    Email: \{firstname.lastname\}@sydney.edu.au, 
    a.pothpitiyage$\_$don@unsw.edu.au, sajal@catharsis.net.au
}}

% \author{\IEEEauthorblockN{Yasod Ginige}
% \IEEEauthorblockA{\textit{School of Computer Science} \\
% \textit{University of Sydney}\\
% Sydney, Australia \\
% yasod.ginige@sydney.edu.au}
% \and
% \IEEEauthorblockN{Akila Niroshan}
% \IEEEauthorblockA{\textit{School of Electrical and Telecommunication Engineering} \\
% \textit{University of New South Wales}\\
% Sydney, Australia \\
% a.pothpitiyage$\_$don@unsw.edu.au}
% \and
% \IEEEauthorblockN{Sajal Jain}
% \IEEEauthorblockA{\textit{Catharsis Pvt. Ltd.} \\
% Sydney, Australia \\
% sajal@catharsis.net.au}
% \and
% \IEEEauthorblockN{Suranga Seneviratne}
% \IEEEauthorblockA{\textit{School of Computer Science} \\
% \textit{University of Sydney}\\
% Sydney, Australia \\
% suranga.seneviratne@sydney.edu.au}
% }

\maketitle

\begin{abstract}
Penetration testing and vulnerability assessment are essential industry practices for safeguarding computer systems. As cyber threats grow in scale and complexity, the demand for pentesting has surged, surpassing the capacity of human professionals to meet it effectively. With advances in AI, particularly Large Language Models (LLMs), there have been attempts to automate the pentesting process. However, existing tools such as PentestGPT are still semi-manual, requiring significant professional human interaction to conduct pentests. To this end, we propose a novel LLM agent-based framework, AutoPentester, which automates the pentesting process. Given a target IP, AutoPentester automatically conducts pentesting steps using common security tools in an iterative process. It can dynamically generate attack strategies based on the tool outputs from the previous iteration, mimicking the human pentester approach. We evaluate AutoPentester using Hack The Box and custom-made VMs, comparing the results with the state-of-the-art PentestGPT. Results show that  AutoPentester achieves a 27.0\% better subtask completion rate and 39.5\% more vulnerability coverage with fewer steps. Most importantly, it requires significantly fewer human interactions and interventions compared to PentestGPT. Furthermore, we recruit a group of security industry professional volunteers for a user survey and perform a qualitative analysis to evaluate AutoPentester against industry practices and compare it with PentestGPT. On average, AutoPentester received a score of 3.93 out of 5 based on user reviews, which was 19.8\% higher than PentestGPT.\\
Code: \textcolor{blue}{https://github.com/YasodGinige/AutoPentester}\\
\end{abstract}


\section{Introduction}

\begin{figure*}[t]
    \centering
    \includegraphics[width=0.95\linewidth]{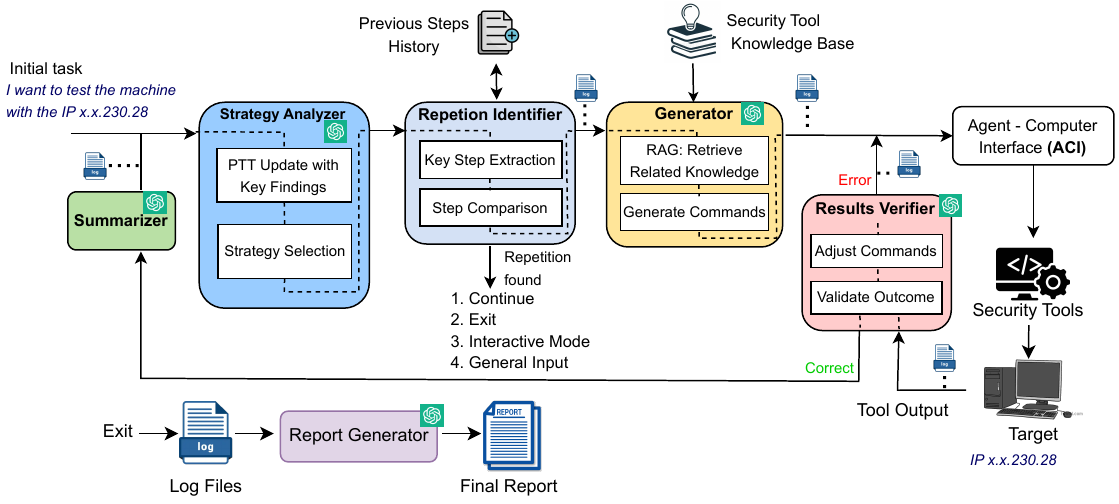}
    \caption{AutoPentester Framework (LLM icons indicate separate API sessions with an LLM).}
    \label{fig: main}
\end{figure*}

Cyber incidents and attacks, whether they are data breaches, ransomware, espionage, phishing, or business email compromises, are increasing at an alarming rate globally~\cite{IBM}. The attackers no longer target only larger and wealthier corporations. Instead, they now focus on SMEs~\cite{SME_cyber}, public sectors, and even essential services such as hospitals~\cite{rahim2024cybersecurity} and emergency services, in addition to individuals. This has created an environment where every corporate computer system, irrespective of its scale, needs to be secured and maintain a good security posture. The cybersecurity industry is unable to meet this demand for skilled professionals~\cite{crumpler2022cybersecurity}.

% Strong regulatory recommendations sometimes mandate these tasks at specific minimum intervals.

Penetration testing (or pentesting) and vulnerability and threat assessment are two essential routine security measures in protecting software and networked systems. \textit{Pentesting} involves simulating real-world attacks on an organization's systems to identify security weaknesses before malicious actors can exploit them. \textit{Vulnerability and threat assessment}, on the other hand, focuses on systematically identifying, analyzing, and prioritizing security risks to help organizations mitigate potential threats proactively. The frequency of pentesting and vulnerability and threat assessments is often governed by mandatory regulations, which vary depending on the industry or government sector~\cite{cis_2024, fedramp_2024}.

Despite their significance, pentesting and vulnerability and threat assessments are repetitive and time-consuming processes that require significant manual effort and expertise~\cite{al2018study}. While the duration varies by task, a typical penetration test can take over two weeks—a considerable timeframe given the growing demand for security assessments~\cite{engebretson2013basics}. Consequently, automating these processes is crucial for an industry already facing a critical shortage of professionals.

Early attempts to automate the penetration testing (pentesting) process leveraged Reinforcement Learning (RL) methods~\cite{hu2020automated, tran2021deep}. However, these approaches primarily focused on suggesting attack vectors rather than executing them or generating comprehensive reports. More recent work~\cite{PentestGPT, Penheal, AutoAttacker, shen2024pentestagent} explored using LLMs or LLM-based agents for automated pentesting.  However, they have several limitations: \textit{\textbf{(a)} their strategy identification is limited without human guidance, leading to trivial or repetitive attack strategies, without making effective progress, \textbf{(b)} the level of automation remains low, as human intervention is often required to execute commands and process outputs on behalf of the tool, \textbf{(c)} they primarily function as advisory systems, offering only basic instructions and handling only a limited number of tools.} Consequently, these solutions have limited effectiveness in automating the pentesting process.

To this end, we propose \textit{AutoPentester}, a large language model (LLM) agent-based framework that delivers a significantly higher level of automation, efficiency, and accuracy across the entire pentesting pipeline. Unlike prior approaches that rely on rigid templates, human guidance, or limited tool handling capabilities~\cite{PentestGPT, AutoAttacker}, AutoPentester introduces a novel architecture composed of five key modules that address the mentioned challenges (a-d):
i) The Strategy Analyzer analyzes the previous steps and their findings to reason out a strategy using Chain-of-Thoughts (CoT), deriving accurate strategies \textbf{(a)}, ii) RAG (Retrieval-Augmented Generation)-based Generator - ensures the creation of advance, accurate and complete commands for multiple tools~\textbf{(c)}, iii) Agent-Computer Interface (ACI) - handles command line based cybersecurity tools to executes these commands~\textbf{(b)}; and iv) Results Verifier - validates outputs and makes necessary adjustments to commands, adding flexibility to the framework and addressing \textbf{(b)}; v) Repetition Identifier - prevents looping issues, and increases the efficiency of AutoPentester~\textbf{(a)}.

Given a target IP address, AutoPentester performs reconnaissance, scanning, vulnerability assessment, and exploitation using multiple cybersecurity tools, and then generates a comprehensive report. We evaluate AutoPentester's strategic penetration testing abilities using Hack-The-Box (HTB) machines and assess its threat analysis capabilities with custom-built vulnerable virtual machines. More specifically, we make the following contributions.

\begin{itemize}
    \item We propose AutoPentester, a novel LLM-agent framework for automated penetration testing, software vulnerability assessment, and threat analysis. AutoPentester overcomes key limitations of existing approaches, such as limited strategic planning, lack of self-adjustments to align with a chosen plan, limited automation, and heavy reliance on human expertise.
    %\item Propose a novel multi agent framework, which can handle multiple security tools, for automated pentesting. The ACI proposed allows the LLM agents to interact with wide range of security tools easily.
    
    \item We evaluate AutoPentester using three LLM backbones and over multiple pentesting and threat assessment tasks, and show that AutoPentester has significant improvements; such as 27.0\% higher subtask completion rate and 39.5\% higher vulnerability coverage compared to the state-of-the-art PentestGPT baseline. % in an automated setting.

    % \item \textcolor{blue}{We evaluate the strategic pentesting ability using HTB machines and threat analysis capability using  of AutoPentester using three LLM backbones and over multiple pentesting and threat assessment tasks and show that AutoPentester has a 12.75\% higher subtask completion rate and 27.78\% higher vulnerability coverage compared to the state-of-the-art PentestGPT. % in an automated setting.}
    
    \item We provide insights into how the various modules of AutoPentester address the limitations of PentestGPT and present an ablation study to demonstrate how these modules work together to generate the pentesting results and reports. Specifically, we show that the RAG module improves subtask completion by 20.0\%, the Repetition Identifier reduces looping by 90.5\%, and the Results Verifier decreases incomplete commands by 80.1\%.
    %\item We create a new benchmark to assess LLM agents in automated pentesting and and show that our framework outperforms PentestGPT by \%.
    \item Finally, we conduct a survey among professional penetration testers and cybersecurity industry experts to evaluate AutoPentester’s performance, the quality of its reports, and how it compares to PentestGPT. AutoPentester achieved an average score of 3.93/5 for all the questions, having a 19.8\% advantage over PentestGPT. Furthermore, participants believe AutoPentester can save time in initial pentesting processes and suggest using it in red teaming tasks.
    
\end{itemize}

The rest of the paper is organized as follows. Section~\ref{related_work} covers related work, while Section~\ref{sec:methodology} introduces the AutoPentester framework. The experimental setup and results are presented in Sections~\ref{sec:experiments} and Section~\ref{sec:results}, respectively. Section~\ref{sec:discussion} discusses the limitations of this work, outlines future directions, and concludes the paper.
\section{Related Work}\label{related_work}

Existing work on automated pentesting can be categorized into two: machine learning-based approaches and LLM-based approaches. \\ 

%We describe related work under two categories; early work that used machine learning-based methods and recent works that have used LLM-based methods.

\noindent{\bf Machine Learning-based methods} Hu et al.~\cite{hu2020automated} proposed a Deep Q-learning-based, two-stage approach for automated penetration testing. It first generates an attack tree using network topology information and uses MulVAL~\cite{ou2005mulval} to identify all possible attack paths. Next, a Deep Q-Learning Network (DQN) is applied to determine the most easily exploitable attack path. HA-DRL~\cite{tran2021deep}, a follow-up work, proposes algebraic action decomposition to manage large discrete action spaces in autonomous penetration testing, achieving faster and more stable optimal attack policies. However, neither method performs actual penetration testing by exploiting software vulnerabilities; instead, they only suggest the best attack vector. NIG-AP~\cite{zhou2019nig} autonomously discovers attack paths of a network by modeling penetration testing as a Markov decision process and using network information for reward-based guidance. However, it does not include software vulnerability assessments and is focused only on network information gathering. Similarly, Casola et al. \cite{casola2018towards} propose a penetration testing method tailored for cloud applications, but it relies on prior knowledge of the application's architecture and security-related data, which must be accurately correlated with the target application. Overall, these works do not cover the complete pentesting scope; rather, they aim to identify the optimal attack vector in specific settings. \\ 

%. They either aim only to identify the optimal attack vector within a network or concentrate on specific aspects of penetration testing, such as cloud applications, within a white-box context.   %\textcolor{orange}{Therefore, despite the advances, these limitations underscore the necessity for further advancements in AI techniques to improve the efficiency and scalability of automated penetration testing tools.}\\

\noindent{\bf LLM-based methods} Recently, LLMs such as GPT~\cite{brown2020language}, Llama~\cite{touvron2023llama}, and Gemini~\cite{team2023gemini} have established new standards across various natural language processing tasks. Since these models were pre-trained on large volumes of internet data, they have substantial knowledge of software vulnerabilities, 
 cybersecurity tools, and vulnerability exploitation, allowing researchers to use them in security-related tasks such as traffic monitoring, intrusion detection, vulnerability analysis, and pentesting.~\cite{trafficgpt, hu2024llm, AutoAttacker}.

ScriptKiddie by Moskal et al. ~\cite{scriptkiddie} is one of the attempts to automate cybersecurity tasks using LLM agent systems. The framework was designed to pursue specific tasks, such as exfiltrating an email server, which can be a subtask of a penetration test. Furthermore, it only generates high-level steps to be followed, and a human must complete the steps and execute them manually. PentestGPT~\cite{PentestGPT} is the first major work towards automating pentesting. It proposes an LLM-based framework to observe a target machine and develop a strategic plan dynamically, using three main modules: summarizer, analyzer, and generator. Each module queries an LLM to complete the assigned work. Given the IP address of the target machine, the analyzer builds a strategic attack path and finds the next step to take, and the generator generates commands and instructions to complete the selected task. The authors tested PentestGPT on ten vulnerable machines in Hack-The-Box (HTB)~\cite{hackthebox_2024}. PentestGPT has two key limitations. First, generated commands need to be executed by a human operator using security tools and report the results back.   Second, to perform some proposed actions, the operator needs to have significant security expertise and conduct their own research.

%Therefore, this process is not fully automated, requiring a considerable amount of human effort.

%\textcolor{red}{are we calling pentestgpt multi agent?/Also the text doesn't explain how it uses an LLM. Add something on where they tested/experiment settings and say that notable human involvment and expertise.}

 AutoAttacker~\cite{AutoAttacker} proposes a post-breach attack framework with a multi-agent structure similar to PentestGPT. Additionally, it has integrated Metasploit tool into the framework such that the generated commands can be directly executed on Metasploit, making the pentesting process automated. However, this work is limited to the Metasploit framework, which is not sufficient for realistic pentesting. Furthermore, it lacks a satisfactory evaluation since the evaluation is done only on the well-known Metasploitable II~\cite{vulnhub} machine. Finally, PenHeal~\cite{Penheal} has comparable limitations.
 
 %proposes a dual module framework for vulnerability identification and remediation. However, it can handle only two security tools and has been tested on only the Metasploitable II machine. 
 
 %Therefore, it lacks a substantial evaluation process. Furthermore, neither of the above tools has been tested in industrial environments, nor have they done any user study using industry professionals to evaluate how suitable their tools are for the industry.

Overall, existing frameworks offer limited automation, cater to specific use cases, and require significant human expertise and intervention. Furthermore, they lack pentesting report generation, a critical real-world need, and remain untested by industry professionals, their ultimate target users.

\section{AutoPentester Framework}
\label{sec:methodology}

AutoPentester is designed to replicate the human approach to penetration testing and assist professionals in pentesting by managing time-consuming groundwork. A human pentester begins with reconnaissance and progressively exploits vulnerabilities based on insights gained from previous steps. Throughout the process, the pentester analyzes results, refines their strategy, selects the most effective next action, and executes it. This iterative cycle continues until all services and software have been assessed. Following the same principle, our framework operates iteratively, executing one task per iteration and dynamically adapting its attack strategy based on prior outcomes.

In AutoPentester, this target behavior is implemented using five LLM-based agents: \textit{Summarizer}, \textit{Strategy Analyzer}, \textit{Generator}, \textit{Results Verifier}, and \textit{Report Generator} as illustrated in Figure~\ref{fig: main}. The Summarizer interprets the tool outputs to a human-readable format, the Strategy Analyzer observes the attack environment and plans the attack path, and the Generator generates suitable commands to execute a selected step. Given an initial task, i.e., an IP address of a target machine, AutoPentester divides the task into subtasks and executes them iteratively. Generally, it begins with an nmap scan, identifying open ports and services for vulnerability assessment, and then moves to each service for vulnerability assessment and exploitation. 

Next, we present the agents and modules of AutoPentester.

\subsection{Summarizer Agent}\label{sec:summarizer}

The input to the Summarizer is the tool output from the previous step, which is the tool output verified by the Results Verifier. Here, a tool can be any common security tool that can be operated through the Command Line Interface (CLI), such as Nmap, Nikto, Metasploit, Dirbuster, and curl. These outputs are usually lengthy; therefore, they often exceed the input token limit of LLMs. To address this, the Summarizer divides the tool output into 6,000-character chunks, each with a 500-character overlap to ensure contextual continuity between them, and then uses an LLM to summarize each chunk. 
%To address this, the Summarizer divides the tool output into 6,000-word chunks, \textcolor{blue}{having 500 words overlapping to build the contextual connection between chunks}, and gets each chunk summarized by an LLM. 
Finally, all the summaries are merged together, again using the LLM to generate the final output. This allows the Strategy Analyzer to easily interpret the previous results and update the attack environment.

\subsection{Strategy Analyzer Agent}\label{sec:analyzer_ag}

\begin{figure}[t]
    \centering
    \includegraphics[width=1\linewidth]{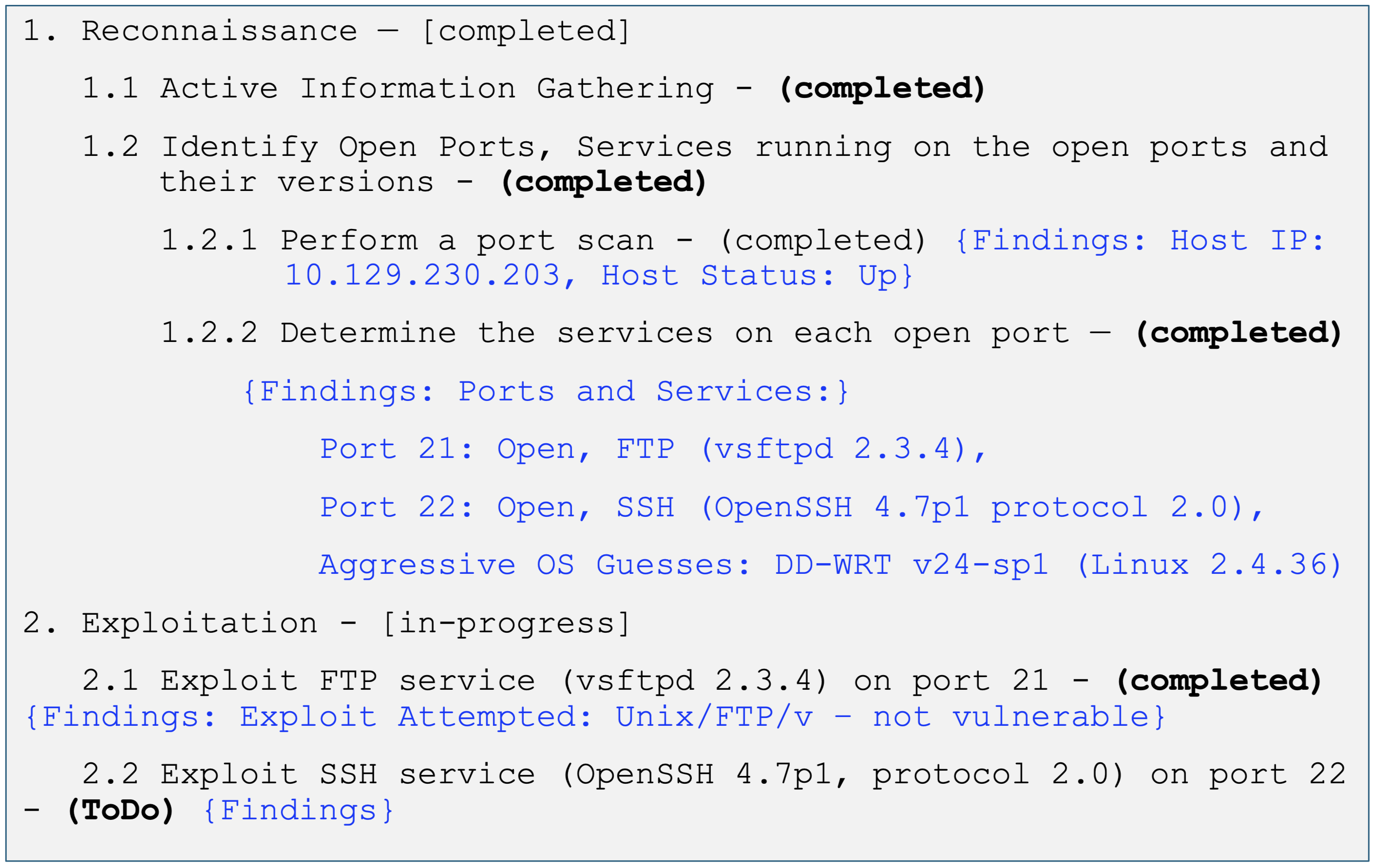}
    \caption{An Example partial PTT (findings in step are progressively added as attributes).}
    \label{fig: PTT}
\end{figure}

The Strategy Analyzer is the strategic planner of AutoPentester. 
%\textcolor{blue}{As shown in Figure~\ref{fig: PTT}, we introduce a modified attack tree structure (Pentest Tree - PTT) compared to PentestGPT~\cite{PentestGPT}, which stores both steps in pentesting workflow and the key findings of each step as attributes.}
It is inspired by PentestGPT, with the major difference of a modified attack tree structure (Pentest Tree - PTT) and the findings-oriented CoT reasoning process to derive the current attack strategy. In the PTT, we store both steps in the pentesting workflow and the key findings of each step as attributes.
For example, Subtask 1.2.2 of Figure~\ref{fig: PTT} contains the discovered information of open ports and the services using nmap. This allows Strategy Analyzer to derive logical and advanced strategies through findings-oriented reasoning.

More specifically, the process of the Strategy Analyzer is twofold. First, it updates the PTT based on the summarized results of the previous step given by the Summarizer. Secondly, it analyzes the gathered information in the PTT with its previous strategy and, through chain-of-thought reasoning, formulates an updated strategy. It then determines the next best step based on this reasoning. For example, as in Figure~\ref{fig: PTT}, following a port scan that identifies vsftpd 2.3.4 running on the FTP port, the logical strategy is to perform a vulnerability assessment on that specific version. The next step would be to use Metasploit to search for available exploits and test them on the target machine. When updating the strategy, we fine-tuned the prompt to check the alignment of new information with the previous strategy and steps, which are in the session context of the LLM, and then select an incremental step in the current subtask, with the target of obtaining root access eventually. If a particular subtask is completed, it will select a new subtask, marked \textit{TODO} in the PTT. This findings-based reasoning process leads to better strategies compared to PentestGPT, which will be further discussed in Section~\ref{sec:results}. 

Furthermore, the Strategy Analyzer tends to forget the PTT when the process runs over a larger number of iterations. To avoid this, we store the PTT in \textit{\texttt{txt}} format and feed it to the Strategy Analyzer as content at each iteration. This draws more attention to the PTT and allows picking the correct steps whilst keeping the focus on strategy. It also enables running the pentest for a larger number of iterations without hallucinating and forgetting the attack environment, which is a major limitation in other work. An example of Strategy Analyzer's output is later given in Figure~\ref{fig: examples}(a).
%\textcolor{red}{is there anything you do differently from PentestGPT - in that case, make it clear.}

% \begin{figure}[h!]
%     \centering
%     \includegraphics[width=1\linewidth]{./Figures/examples.png}
%     \caption{Examples for each module.}
%     \label{fig: examples}
% \end{figure}

\subsection{Repetition Identifier Module}\label{rep_identifier}

Occasionally, if a particular subtask fails in the exploitation, the framework tends to get stuck in it by trying similar methods in a loop. For example, it will keep trying the same Metasploit exploitation even if the service is not vulnerable to it. To avoid this, we introduce a Repetition Identifier module that checks the similarity of the current step with the previous steps (in previous iterations) of the process. Using the Strategy Analyzer's output, we structure the selected step to briefly describe what service will be exploited, how it is done, and what tool it uses. Then, we generate a vector embedding for that description and store it separately. When a new step is given, we check whether there is a similar task that has been executed before. Here, we use the cosine similarity score and a manual threshold of 0.15 (chosen empirically), where we consider values below that as repetitions. In case of a repetition, four options are given to the operator, while the rest is passed to the Generator agent.

\begin{enumerate}
    \item \textbf{Continue:} If the operator doesn't give any input, the program will continue to try a different path.
    \item \textbf{Exit:} Exit the iterative process and generate the report with the collected information.
    \item \textbf{Interactive Mode:} The framework provides an interactive mode for GUI tools such as Burp Suite~\cite{burpsuite}, i.e., the operator manually executes commands based on the generator’s instructions and provides text feedback to AutoPentester.
    \item \textbf{General Input:} The user can give a general instruction to the framework. The program will be pointed back to the summarizer with the given instructions.
    
\end{enumerate}

An example of the functionality of the Repetition Identifier is given in Figure~\ref{fig: examples}(e).

\subsection{Generator Agent}

\begin{figure}[ht]
    \centering
    \includegraphics[width=0.62\linewidth]{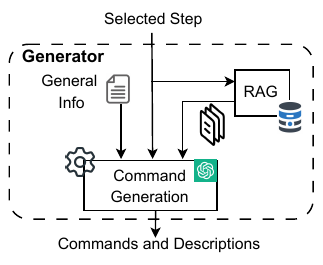}
    \caption{The functionality of the Generator Agent.}
    \label{fig: generator}
\end{figure}

Given the best step to be taken in the current iteration recommended by the Strategy Analyzer, the Generator agent generates the commands that can be run on CLI-based pentesting tools such as Nmap, Metasploit, and Dirbuster. Even though current LLMs have substantial knowledge of cybersecurity tasks, we observed that they tend to hallucinate when they don't have knowledge about a particular attack using a particular tool. To solve this, we use a RAG (Retrieval-Augmented Generation) architecture, which retrieves the relevant knowledge for the selected step from a knowledge base as illustrated in Figure~\ref{fig: generator}. Furthermore, we feed the general information about the attack machine (where AutoPentester is running), such as paths to the documents that are needed by tools (e.g., common username and password lists) and the local IP address. This enables the Generator to generate complete and executable commands for various tools without any human input, making a major advancement in automation.\\

\noindent{\bf RAG Module} We use the following resources to create the knowledgebase; \textit{i) Metasploit: The Penetration Tester's Guide~\cite{kennedy2011metasploit}, ii) Penetration Testing: A Hands-on Introduction to Hacking~\cite{weidman2014penetration}, iii) Articles collected from the HackTrics platform~\cite{hacktricks}.} We selected the first two resources as suggested in~\cite{Penheal}. As they do not encompass all common tools, such as Dirbuster and Nikto, we gathered a collection of online articles that include relevant commands and explanations.

% \begin{itemize}
%     \item Metasploit: The Penetration Tester's Guide~\cite{kennedy2011metasploit}
%     \item{Penetration Testing: A Hands-on Introduction to Hacking~\cite{weidman2014penetration}}
%     \item{Articles collected from the HackTrics platform~\cite{hacktricks}.}
% \end{itemize}

To form the database, we preprocessed the text and stored it as vectors. First, the text is divided into 500-character chunks and generates vector embeddings using OpenAI’s text-embedding model, text-embedding-ada-002~\cite{openai}. The resultant vector embeddings were stored in a vector database. When a query (the selected step from the Strategy Analyzer) is received, it is embedded in the same manner, and cosine similarity is used to identify the ten most relevant data chunks. These retrieved chunks are then provided to the Generator as supporting information for command generation. The final outcome of the Generator is a set of commands with instructions to fulfill the selected task by the Strategy Analyzer, as shown in Figure~\ref{fig: examples}(b).

\subsection{Agent Computer Interface (ACI)}

\begin{figure}[ht]
    \centering
    \includegraphics[width=0.97\linewidth]{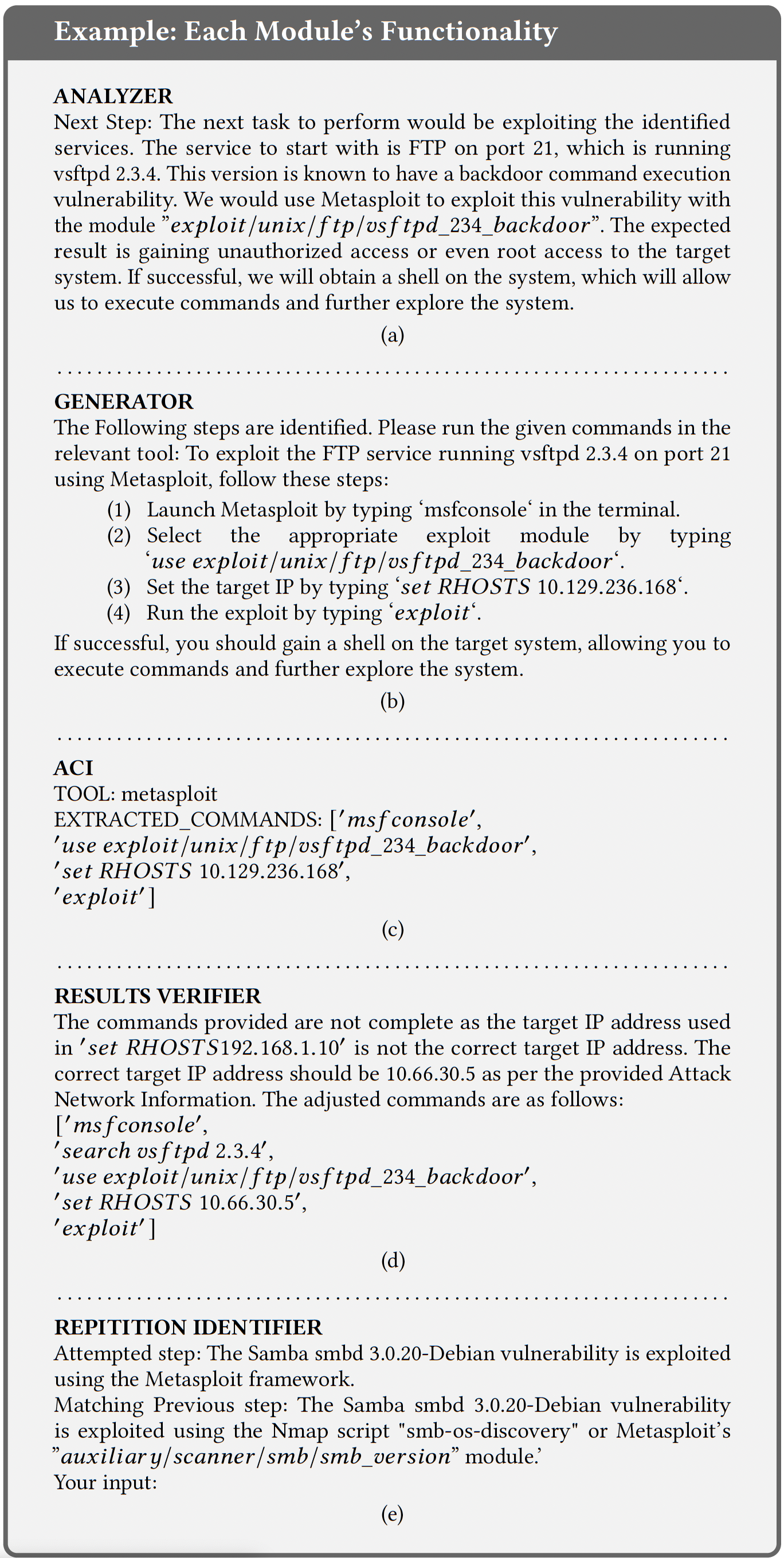}
    \caption{ Examples for each module’s functionality}
    \label{fig: examples}
\end{figure}

The ACI performs two tasks. First, it extracts the commands from the Generator's output using an LLM query and arranges them according to the relevant tool input as shown in Figure~\ref{fig: examples}(c). Then it triggers the relevant Python function (according to the tool), which takes the extracted commands as input arguments. We use the \textit{\texttt{subprocess}}~\cite{subprocess} library to handle general CLI tools such as nmap, dirbuster, and the \textit{\texttt{pexpect}}~\cite{pexpect} library to handle interactive CLIs like Metasploit. The design of ACI is general so that it can be extended to any security tool running on a static or interactive CLI, enabling AutoPentester to use a wide range of cybersecurity tools. The results of the tool will be captured and passed to the results verifier to check the validity. Since neither the Repetition Identifier nor ACI relies on LLM API calls, we do not classify them as agents.

%We have tested it using a set of security tools (Nmap, Metasploit etc.) during our experiments, which are mentioned in the Section~\ref{experiments}. 

%The design of ACI is general so that it can be extended to any security tool running on a static or interactive CLI.
%After selecting the tool, the ACI triggers the corresponding function, giving the extracted commands as inputs. %An example of command extraction is given in Figure~\ref{fig: examples}(c).

%\subsection{Security tools}

\subsection{Results Verifier Agent}

The output of the security tool may not be the expected outcome at the very first attempt. For example, if the ports are filtered on the target machine, the nmap commands without \texttt{-Pn} or \texttt{-sS} would not reveal the open ports. In such situations, we use the Results Verifier to check the tool outputs and refine the commands (generated by the Generator) using reasoning to obtain better outcomes. The inputs of the Results Verifier are the commands generated by the Generator and the outcomes of the security tools. If the outcomes of the security tools are correct, they are directed to the Summarizer. An example of the functionality of the Results Verifier is given in Figure~\ref{fig: examples}(d).

\subsection{Report Generator Agent}

We maintain a log file to record all the information gathered throughout the pentesting process in a human-readable format. When the pentesting process is complete \textbf{(Exit)}, the Report Generator goes through the log file and creates a \texttt{.csv} file summarising the findings, which is a widely used practice in the industry for pentesting reports. It contains the following vulnerability information: CVE number, CVSS score, Risk level, Protocol, Port, Vulnerability name, Synopsis, Description, Solution (Remediation), Hostname, IP address, OS, Reference URL, Vulnerability Priority Rating (VPR). This allows users to get a quick overview of the security posture of their system. Furthermore, the log file provides a detailed description of the process.

%We have published our code on a GitHub repository. \footnote {https://github.com/YasodGinige/AutoPentester}
\section{Experiment Setup}\label{sec:experiments}

We set up the AutoPentester framework in a Linux virtual machine and installed the following security tools on it: Nmap, Metasploit, Netcat, Nikto, Dirbuster, John the Ripper, Sqlmap, Smbclient, Dnsrecon, and SSLscan. We selected these specific tools based on previous work~\cite{PentestGPT, AutoAttacker} and domain knowledge on the most commonly used pentesting tools. The ACI can use these tools to run commands and test the target machine. %\textcolor{blue}{how to ensure the safety of the process?? Fully automated functionality ....}

\subsection{Target Machines}\label{test_bench}

% \begin{table}[b]
% \footnotesize
% \centering
% \caption{VMs containing OWSAP Top-10 vulnerabilities}
% \begin{tabular}{lcp{4.5cm}c}
% \hline
% \textbf{Name} & \textbf{Num. Vul.} & \textbf{Vulnerabilities} & \textbf{Ports} \\
% \hline
% VM1 & 4 &
% vsftpd 2.3.4 backdoor attack~\cite{cve_vstpd} & 21 \\
% & & OpenSSH username enumeration~\cite{cve_openssh} & 22 \\
% & & Samba remote code execution~\cite{cve_samba} & 139 \\
% & & Telnet remote code execution~\cite{cve_telnet} & 23 \\
% \hline

% VM2 & 4 & vsftpd 2.3.4 backdoor attack~\cite{cve_vstpd} & 21 \\
% & & Smtp remote code execution~\cite{cve_smtp} & 25 \\
% & & Telnet remote code execution~\cite{cve_telnet} & 23 \\
% & & HTTP server default credentials & 8888 \\
% \hline

% VM3 & 4 & WordPress SQL injection~\cite{cve_wordpress} & 80 \\
% & & OpenSSH username enumeration~\cite{cve_openssh} & 22 \\
% & & Smtp remote code execution~\cite{cve_smtp} & 23
% \\
% & & Samba remote code execution~\cite{cve_samba} & 139\\

% \hline
% VM4 & 9 & Metasploitable II vulnerabilities~\cite{vulnhub} & x \\
% \hline
% \end{tabular}

% \label{tab:VM_list}
% \end{table}

We use two sets of target machines to assess AutoPentester. \\ 

\noindent{\bf Hack the Box Machines} Hack the Box (HTB)~\cite{hackthebox_2024} is an industry-grade pentesting training platform that provides vulnerable machines of varying complexity levels. Each machine contains a set of vulnerabilities, and users have to exploit them strategically while connecting the information revealed to gain root access. We use these machines to measure the strategic pentesting ability of AutoPentester. We selected 10 machines from Hack the Box, consisting of six easy and four intermediate difficulty levels, that contain different vulnerabilities, which lead to single or multiple attack vectors. We selected five machines used in the PentestGPT study (Sau, Pilgrimage, Topology, Authority, and Jupiter) as benchmark references and included five additional machines to ensure a fair comparison for both tools on unseen machines. The chosen machines span a broad spectrum of subtasks commonly encountered in practical pentesting. A detailed breakdown of these subtasks is presented in Table~\ref{tab:subtasks}. We used the official write-ups given for each machine on the HTB platform to extract those subtasks which can be categorised into nine categories: Port scanning, Web enumeration, Network enumeration, Crypto analysis, File enumeration, Shell construction, Command injection, Source code analysis, and Known attacks, such as attacks listed in GitHub for specific vulnerabilities.

% \textcolor{blue}{
% \begin{itemize}
%     \item Port scanning: includes scans searching for open ports and services. The most common approach is the nmap scan.
%     \item Web enumeration: includes tasks related to web applications such as default credential login and cookie capturing.
%     \item Network enumeration: includes tasks related to networking related services such as SSH, FTP, SMTP, and Active Directory.
%     \item Crypto analysis: includes decryption tasks of revealed encrypted information.
%     \item File enumeration: includes tasks related to analyzing the content of the files and extracting important information.
%     \item Shell construction: includes tasks related to reverse shell constructions.
%     \item Command injection: includes tasks related to command injections such as SQL injection, payload injection, and cross-site scripting.
%     \item Known exploits: includes tasks related to using known exploits for specific vulnerabilities. This includes cloning a GitHub repository and executing an exploitation.
%     \item Code analysis: includes tasks related to source code analysis.
% \end{itemize}}

%Users have to exploit these vulnerabilities and strategically find a correct attack vector to hack into the machine and obtain the flag. 

\begin{table}[t]
\footnotesize
\centering
\caption{Vulnerability selection covering OWASP top 10. \footnotesize The Metasploitable II machine is used as VM4.} 
\begin{tabular}{llp{0.4cm}p{0.4cm}p{0.4cm}}
\hline
\textbf{Vulnerability} & \textbf{OWASP} & \textbf{VM1} & \textbf{VM2} & \textbf{VM3} \\
\hline
vsftpd 2.3.4 backdoor attack~\cite{cve_vstpd} & 4,6,7 & \checkmark & \checkmark &  \\
OpenSSH username enumeration~\cite{cve_openssh} & 1,7.9 & \checkmark &  & \checkmark \\
Samba remote code execution~\cite{cve_samba} & 2,4,5 & \checkmark &  & \checkmark \\
Telnet remote code execution~\cite{cve_telnet} & 3,4 & \checkmark & \checkmark &  \\
SMTP remote code execution~\cite{cve_smtp} & 3,8 &  & \checkmark & \checkmark \\
HTTP server default credentials & 1,4 &  & \checkmark &  \\
WordPress SQL injection~\cite{cve_wordpress} & 1,3 &  &  & \checkmark \\
\hline
\label{tab:VM_list}
\end{tabular}
\end{table}

Furthermore, we did not include any machine we used for prompt fine-tuning (Bank, Forest, and Bike) and functionality checking during the development phase among these targets.\\

\noindent{\bf Custom VMs} In addition to HTB machines, we also created four custom virtual machines containing vulnerabilities in OWASP top-10~\cite{owasp_2021}. In contrast to HTB machines, which are used to measure AutoPentester's strategic pentesting ability, these machines were specifically designed to evaluate its performance in vulnerability and threat assessment. Here, the main focus is on maximizing vulnerability identification and achieving a higher coverage of the attack surface. We built these machines in VirtualBox~\cite{oracle_2023} and planted the vulnerabilities by installing vulnerable software versions and applications. For VM4, we directly used the Metasploitable II VM in VirtualBox. The details of these machines are given in Table~\ref{tab:VM_list}.

\subsection{PentestGPT Baseline}

%We evaluate our tool both quantitatively and qualitatively, comparing it to PentestGPT—the only similar work with a reproducible repository.\\ 

We chose PentestGPT~\cite{PentestGPT} as our baseline because it serves a similar purpose and has publicly available code. Other related works, such as AutoAttacker~\cite{AutoAttacker} and Penheal~\cite{Penheal}, lack public code, preventing reproducibility. When selecting target machines, we ensured that they contain five machines from the PentestGPT baseline experiments as mentioned in Section~\ref{test_bench}.

In addition to HTB, PentestGPT was also tested on picoCTF challenges~\cite{picoctf_2021}. However, we observed that the operators must provide critical information about the correct attack vector to achieve the reported results. For instance, the XtraORdinary challenge involves a random iterative encryption. Although the paper claims it was solved, PentestGPT was unable to generate a solution without human assistance, such as providing significant guidance on the approach and task breakdown. Therefore, we omitted the picoCTF testing from our experiments. 

%Additionally, PentestGPT has not conducted a user study with industry experts to assess its usability.

%These 10 machines 

\begin{table*}[t]
\footnotesize
    \centering
    \caption{\centering{HTB machines used in performance evaluation (L-Linux, W-Windows). (*) represents the number of services or subtasks.}}
    \begin{tabular}{lcp{5cm}p{8.5cm}}
        \hline
        \textbf{Machine} & \textbf{Level} & \textbf{Services (Ports)} & \textbf{(Num. of Sub tasks) - Sub task order} \\
        \hline
        Lame (L) & Easy & \textbf{(3)} - FTP: 21, SSH: 22, netbios-ssn: 139/445  & \textbf{(4)} - Port scanning \texttt{->} Network Enumeration \texttt{->} Known Vuln \texttt{->} Shell creation \\
        \hline
        Bashed (L) & Easy & \textbf{(1)} - HTTP: 80 & \textbf{(4)} - Port scanning \texttt{->} Web Enumeration \texttt{->} File Enumeration \texttt{->} Shell Construction \\
        \hline
        Active (W) & Easy & \textbf{(9)} - DNS: 53, Kerberos: 88, RPC: 135, netbios-ssn: 139, LDAP: 389, microsoft-ds: 445, kpasswd5: 464, RPC: 593 & \textbf{(6)} - Port scanning \texttt{->} Network enumeration \texttt{->} File enumeration \texttt{->} Cryptoanalysis \texttt{->} Network enumeration \texttt{->} File enumeration \\
        \hline
        Sau (L) & Easy & \textbf{(3)} - SSH: 22, HTTP: 80, Unknown: 55555 & \textbf{(6)} - Port scanning \texttt{->} Web enumeration \texttt{->} Shell construction \texttt{->} Website enumeration \texttt{->} Known exploits \texttt{->} File enumeration \\
        \hline
        Pilgrimage (L) & Easy & \textbf{(2)} - SSH: 22, HTTP: 80 & \textbf{(7)} - Port scanning \texttt{->} Web enumeration \texttt{->} File enumeration \texttt{->} Code analysis \texttt{->} Known exploit \texttt{->} File enumeration \texttt{->} Cryptoanalysis \\
        \hline
        Topology (L) & Easy & \textbf{(2)} - SSH: 22, HTTP: 80 & \textbf{(7)} - Port scanning \texttt{->} Web enumeration \texttt{->} Command injection \texttt{->} Network enumeration \texttt{->} File enumeration \texttt{->} Command injection \texttt{->} File enumeration \\
        \hline
        Authority (W) & Medium & \textbf{(16)} - DNS: 53, HTTP: 80, Kerberos: 88, MSrpc: 135, netbios-ssn: 139, LDAP: 389, kpasswd5: 464, ncacn-http: 593, SSL: 636 & \textbf{(7)} - Port scanning \texttt{->} Web enumeration \texttt{->} Network enumeration \texttt{->} File enumeration \texttt{->} Code analysis \texttt{->} Cryptoanalysis \texttt{->} Web enumeration \\
        \hline
        Jupiter (L) & Medium & \textbf{(2)} - SSH: 22, HTTP: 80 & \textbf{(7)} - Port scanning \texttt{->} Network enumeration \texttt{->} Web enumeration \texttt{->} Command injection \texttt{->} Shell construction \texttt{->} File enumeration \texttt{->} Network enumeration \\
        \hline
        Ambassador (L) & Medium & \textbf{(4)} - SSH: 22, HTTP: 80, Unknown: 3000, MySQL: 3306 & \textbf{(7)} - Port scan \texttt{->} Web enumeration \texttt{->} Network enumeration \texttt{->} Known exploitation \texttt{->} File enumeration \texttt{->} Command injections \texttt{->} Network enumeration \\
        \hline
        Jarvis (L) & Medium & \textbf{(2)} - SSH: 22, HTTP: 80 & \textbf{(6)} - Port scanning \texttt{->} Web enumeration \texttt{->} Command injection \texttt{->} File enumeration \texttt{->} Shell construction \texttt{->} Code analysis \\
        \hline
    \end{tabular}
    
    \label{tab:subtasks}
\end{table*}

\subsection{Performance Metrics}\label{sec:P_metric}

We evaluate performance using quantitative and qualitative metrics. 

\noindent\textbf{Quantitative analysis:} We use the following metrics to measure a tool's performance quantitatively.

\begin{itemize}
    \item \textbf{Subtask Completion \% ($\uparrow$)} – The percentage of subtask completion for each machine. These subtasks were manually extracted from the official write-ups for each machine, available on the HTB platform. The subtask breakdown is given in Table~\ref{tab:subtasks}.
    \item \textbf{Services Coverage \% ($\uparrow$)} – The number of services (such as HTTP, FTP, and SSH) covered during the pentesting process.
    \item \textbf{No. of Steps  ($\downarrow$)} – The number of steps used for each machine. A step refers to a complete iteration in the tool as illustrated in Figure~\ref{fig: main}.
    \item \textbf{No of. Loops ($\downarrow$)} – The number of repetitive steps where the tool tries the same step repeatedly.
    \item \textbf{Human Interaction ($\downarrow$)} – The number of steps requiring human intervention, including:
    \begin{itemize}
        \item Running commands on a security tool (e.g., Metasploit) and reporting results to the agent.
        \item Adjusting or correcting incomplete commands.
        \item Observing visual results (e.g., in web applications) and interpreting them in text format for the tool.
    \end{itemize}
    \item \textbf{No. of Incomplete Commands ($\downarrow$)} – The number of incomplete or erroneous commands generated by the tool.
    \item \textbf{Vulnerability Coverage ($\uparrow$)} – The number of vulnerabilities covered during the pentesting process. Note that this metric is used only with the custom VM experiments where we know the exact number of existing vulnerabilities.
\end{itemize}

%For task completeness, we measure the sub task completion percentage for each machine. For the attack surface coverage, we measure the number of services covered during the pentesting process. We count the number of steps used for each machine to measure the efficiency. Here, a step mean a complete iteration in the tool as illustrated in Figure~\ref{fig: main}. We count the number of incomplete or erroneous commands generated by the tools to measure the command correctness. To measure looping effects, we count the number of repetitive steps (trying the same step repetitively) in the process. To measure the human interaction level, we count the steps that need human interaction with the tool. We consider the following activities as human interactions: running commands on a tool and reporting the results to the agent, adjusting incomplete commands, observing a visual result (such as in web applications), and interpreting it to the tool in text format.

We carefully analyzed the log files generated on each machine during the experiments and manually counted the aforementioned metrics. Subtask completion was determined by verifying if a subtask's requirements, as detailed in the official write-ups, were met. For instance, if a website contained specific information accessible via a vulnerable endpoint, the subtask was considered complete once that endpoint was identified and the information extracted. To minimize human error, two authors independently counted and cross-verified these results. We run experiments three times for each target machine and report average values 
in Section~\ref{quantitative}.\\ 

\noindent\textbf{Qualitative analysis:} We recruited 10 cybersecurity professional volunteers with more than five years of experience via a LinkedIn post. The participants included seven pentesters and three cybersecurity professionals, all actively working in the field. This user study was approved by the Human Research Ethics Committee of The University of Sydney under the application number 2024/HE001529.

% , forming a pool of seven pentesters and three security experts. All of them were active and was employed as a 

% The survey aimed to qualitatively assess the tools' suitability for real-world pentesting tasks.

% We conducted a user survey by recruiting cybersecurity industry professional volunteers under approval from our organization's human ethics committee. We published a post on LinkedIn and recruited professional pentesters and security experts, resulting in a participant pool of seven professional pentesters and three security professionals. The main goal is to qualitatively evaluate how suitable those tools are for actual pentesting tasks.

We generated pentest reports using PentestGPT and AutoPentester for two Hack The Box (HTB) machines (Active and Bashed) and two virtual machines (VM1 and VM4). Industry experts were then asked to complete a questionnaire based on their evaluations. Additionally, we provided screen recordings of the tools to give insight into their functionality. The questionnaire consists of 13 MCQs and 15 short-answer questions. The questions on HTB machines reports were more towards attack strategies and the maturity of the steps and commands of the tools, while the questions on VMs were towards the vulnerability coverage and quality of reporting. However, both HTB machines and VMs shared four common questions regarding attack surface coverage (Q1), the extent of disclosed information (Q2), the progression of attack steps (Q3), and the efficiency of those steps (Q5). The questions and survey results are given in Section~\ref{qualiitative}. We have uploaded the experiment log files, questionnaire, and responses to our GitHub repository.

\subsection{Choice of LLM} \label{sec:LLMchoice}

AutoPentester can be integrated into any LLM that provides an API.  Different LLMs will have different knowledge levels on cybersecurity-related tasks depending on their training data. Therefore, they may perform differently on pentesting tasks.

To pick the best LLM backend for AutoPentester, we first tested AutoPentester using the three LLMs; Gemini-2.0-flash, GPT-3.5-turbo, and GPT-4-turbo on HTB machines to test their ability to understand the attack environment ({\bf cf.} Table~\ref{tab:model_comp}). We selected the LLM with the highest subtask completion percentage, which was GPT-4-turbo, as the backbone for the rest of the experiments.

% \subsection{Artifacts and Log Files Release}
% \label{sec:Artifacts}

%Once we identified the best-performing model, which was GPT-4 as explained in Section~\ref{sec:LLMResult}, the rest of our experiments used GPT-4 as the LLM backend.

%First we evaluate the capabilities of LLMs in pentesting tasks by using three LLMs (Gemini, GPT-3.5-turbo and GPT-4-turbo) in agents of the AutoPentester and evaluating them on ten Hack The Box (HTB)~\cite{hackthebox_2024} machines. For convenience, we refer them as GPT-3.5 and GPT-4 in following sections. 

\begin{table}[t]
\small
    \centering
    \renewcommand{\arraystretch}{0.9} 
    \caption{Performance comparison of different LLMs. Here (L) represents Linux and (W) represents Windows.}
    \begin{tabular}{llccc}
    \toprule
    \textbf{Name} & \textbf{Level} & \multicolumn{3}{c}{\textbf{Subtask Completion (\%)}} \\
    \cmidrule(lr){3-5}
    & & \textbf{Gemini} & \textbf{GPT-3.5} & \textbf{GPT-4} \\
    \midrule
    Lame (L) & Easy & 50.00 & 100.00 & 100.00 \\
    Bashed (L) & Easy & 50.00 & 58.33 & 100.00 \\
    Active (W) & Easy & 28.57 & 47.62 & 57.34 \\
    Sau. (L) & Easy & 16.67 & 33.33 & 44.44 \\
    Pilgrimage (L) & Easy & 28.57 & 33.33 & 33.33 \\
    Topology (L) & Easy & 14.28 & 23.80 & 52.38 \\
    Authority (W) & Medium & 28.57 & 38.09 & 52.38 \\
    Jupiter (L) & Medium & 25.00 & 29.17 & 33.33 \\
    Ambassador (L) & Medium & 28.57 & 42.85 & 61.90 \\
    Jarvis (L) & Medium & 16.67 & 16.67 & 50.00 \\
    \midrule
    \textbf{Average} & & 28.69 & 42.32 & \textbf{58.51} \\
    \bottomrule
\end{tabular}
    \label{tab:model_comp}
\end{table}

\section{Results}\label{sec:results}

\begin{table*}[t]
\footnotesize
    \centering
    \renewcommand{\arraystretch}{0.9}
    \caption{Comparison of AutoPentester (AP) and PentestGPT (P) Performance on HTB machines. We run experiments three times for each machine and report the average. Note that AutoPentester achieves a higher subtask completion percentage and service coverage with a lower number of steps. Also, it has a significantly lower number of loops, human interaction and incomplete commands compared to PentestGPT.}
    \begin{tabular}{lcp{1.8cm}p{1.7cm}ccp{1.7cm}p{1.9cm}|p{1.2cm}p{0.9cm}}

        \toprule
        \textbf{Name} & \textbf{Tool} & \textbf{Subtask Completion (\%)} ($\uparrow$) & \textbf{Services Covered (\%)} ($\uparrow$) & \textbf{Steps} ($\downarrow$) & \textbf{Loops} ($\downarrow$) & \textbf{Human Interaction} ($\downarrow$) & \textbf{Incomplete Commands} ($\downarrow$) & \textbf{Time (mins)} ($\downarrow$) & \textbf{Cost(\$)} ($\downarrow$)\\
        \midrule
        
        \multirow{2}{*}{Lame} & P & 50.0 & 88.89 & 12.67 & 3.33 & 13.67 & 1.00 & 9.4 & 7.2\\
         & AP & \textbf{100.0} & \textbf{100.0} & \textbf{4.33} & \textbf{0.00} & \textbf{0.00} & \textbf{0.00} & 7.8 & 8.3 \\
         \midrule
         
        \multirow{2}{*}{Bashed} & P & 91.67 & 100.0 & 6.67 & 0.33 & 10.00 & 3.33 & 5.2 & 4.1 \\
         & AP & \textbf{100.0} & \textbf{100.0} & \textbf{5.67} & \textbf{0.00} & \textbf{1.00} & \textbf{0.00} & 15.7 & 9.4 \\
         \midrule
         
        \multirow{2}{*}{Active} & P & \textbf{72.22} & 22.22 & 12.67 & 2.33 & 17.67 & 5.00 & 19.1 & 9.7 \\
         & AP & 66.67 & \textbf{48.15} & \textbf{12.67} & \textbf{0.67} & \textbf{1.67} & \textbf{0.00} & 47.8 & 16.6\\
         \midrule
         
        \multirow{2}{*}{Sau} & P & 33.33 & 100.0 & 14.67 & 3.33 & 18.33 & 3.67 & 29.4 & 15.2\\
         & AP & \textbf{44.44} & \textbf{100.0} & \textbf{12.33} & \textbf{0.33} & \textbf{0.00} & \textbf{0.00} & 45.4 & 15.9 \\
         \midrule
         
        \multirow{2}{*}{Pilgrimage} & P & \textbf{42.86} & 100.0 & 13.00 & 2.33 & 21.00 & 9.67 & 25.8 & 12.6\\
         & AP & 33.33 & \textbf{100.0} & \textbf{9.33} & \textbf{0.67} & \textbf{0.00} & \textbf{0.00} & 37.5 & 13.3 \\
         \midrule
         
        \multirow{2}{*}{Topology} & P & 33.33 & 100.0 & \textbf{12.67} & 2.67 & 16.00 & 3.33 & 25.1 & 9.3 \\
         & AP & \textbf{52.38} & \textbf{100.0} & 13.00 & \textbf{0.00} & \textbf{5.00} & \textbf{0.67} & 48.9 & 16.4 \\
         \midrule
         
        \multirow{2}{*}{Authority} & P & 47.62 & 16.67 & 12.67 & 2.67 & 19.67 & 8.00  & 24.4 & 11.5\\
         & AP & \textbf{52.38} & \textbf{16.67} & \textbf{9.67} & \textbf{0.33} & \textbf{3.00} & \textbf{0.33} & 38.7 & 14.8 \\
         \midrule
         
        \multirow{2}{*}{Jupiter} & P & 33.33 & 100.0 & 9.00 & 2.33 & 17.67 & 8.67 & 20.3 & 6.6 \\
         & AP & \textbf{38.09} & \textbf{100.0} & \textbf{8.67} & \textbf{0.33} & \textbf{0.00} & \textbf{0.00} & 34.4 & 12.2 \\
         \midrule
        \multirow{2}{*}{Ambassador} & P & 28.57 & 25.00 & \textbf{9.33} & 1.00 & 8.67 & 1.00 & 22.5 & 10.6 \\
         & AP & \textbf{61.90} & \textbf{91.67} & 10.67 & \textbf{0.67} & \textbf{0.33} & \textbf{0.33} & 40.3 & 15.1 \\
         \midrule
         
        \multirow{2}{*}{Jarvis} & P & 38.89 & 50.00 & 9.00 & 0.67 & 11.00 & 3.67 & 21.9 & 9.3\\
         & AP & \textbf{50.00} & \textbf{100.0} & \textbf{8.33} & \textbf{0.00} & \textbf{0.33} & \textbf{0.00} & 32.2 & 12.7\\
         \midrule
         
        \multirow{2}{*}{Average} & P & 47.18 & 70.28 & 11.23 & 2.10 & 15.36 & 4.46 & \textbf{20.31} & \textbf{9.61} \\
         & AP & \textbf{59.92} & \textbf{85.64} & \textbf{9.46} & \textbf{0.30} & \textbf{1.13} & \textbf{0.13} & 34.87 & 13.47 \\
        \bottomrule
    \end{tabular}
    
    \label{tab:HTB_comp}
\end{table*}

In this section, we present our results. We first report a performance comparison of AutoPentester against the PentestGPT baseline, followed by the results of the user survey. Finally, we conduct an ablation study to evaluate each module’s contribution to performance.

\subsection{Quantitative Analysis}\label{quantitative}

We evaluate the performance of AutoPentester and PentestGPT on the target machines and performance metrics described in Section~\ref{test_bench} and Section~\ref{sec:P_metric}, respectively. \\ 

\noindent{\bf Hack The Box Machines} In Table~\ref{tab:HTB_comp}, we show the performance results on HTB machines. On average, AutoPentester outperforms PentestGPT in all the metrics. Among all the scenarios in the table, AutoPentester outperforms PentestGPT in 49 cases, matches its performance in seven cases, and underperforms only in four cases.

For instance, AutoPentester has a 27.0\% better average subtask completion rate (59.92\%) compared to PentestGPT (47.18\%). Out of 10 machines, AutoPentester outperformed PentestGPT on 8 machines in subtask completion. Notably, AutoPentester achieves higher performance while requiring approximately 1.7 (18.7\%) fewer steps than PentestGPT. This efficiency is primarily due to PentestGPT’s tendency to get stuck in repetitive loops, repeatedly attempting the same action. In contrast, AutoPentester leverages the Repetition Identifier to detect such loops and shift strategy. %We further discuss this scenario in Appendix~\ref{sec:ap_repe}.

%Furthermore, this higher performance was gained using a lower number of steps (1.53 on average) compared to the PentesGPT. 

This is also reflected in the Loop Count metric, where PentestGPT averages 2.1 loops per machine, while AutoPentester significantly reduces this by 85.7\%, just 0.3 loops per machine. Also, we highlight the automation advantage of Autopentester, which has a significantly lower number of human interactions on average ($1.13$) compared to PentestGPT's 15.36, which was one of our major design goals.

%that PentestGPT has significantly higher human interactions (15.36 per machine on average) compared to the AutoPentester. This indicates that PentestGPT requires human inputs to perform the pentesting tasks.

Finally, the results show that PentestGPT tends to generate more incomplete commands. For example, it suggests commands such as \texttt{`nmap -p- <target IP>'}, \texttt{`smbclient //<target-ip>/<share-\allowbreak name> -N'}, which should be adjusted by a human before running them on a tool. The average number of incomplete commands per machine is 4.46. In contrast, AutoPentester generates complete commands most of the time, having only 0.1 incomplete commands per machine (97.7\% reduction). This can be attributed to the effect of the RAG in the Generator and the actions of the Results Verifier. The RAG provides example commands to help the Generator produce accurate and complete outputs. If an incorrect or incomplete command is generated, the Results Verifier ensures its correction.\\

\noindent{\bf Custom VMs} Next, we present the results on custom VMs. As mentioned in Section~\ref{test_bench}, the objective of this testing is to assess AutoPentester's ability to conduct vulnerability and threat assessments automatically.  As shown in Table~\ref{tab:vuln_cov}, AutoPentester consistently covers a higher fraction of vulnerabilities, achieving 98.14\% on average across all the VMs, while the PetestGPT achieves only 70.37\%. Table~\ref{tab:VM_comparison} gives the step, loop, human interaction, and incomplete command counts for both tools. Note that AutoPentester takes significantly fewer steps to achieve higher vulnerability coverage compared to PentestGPT. This is due to the loops in PentestGPT, which is 3.25\% per VM on average. Furthermore, PentestGPT has a higher human interaction (18.16 per VM) and incomplete command counts (6.25 per VM) compared to almost zero counts for AutoPentester. Overall, the results show that AutoPenster is more effective and efficient in vulnerability and threat assessment tasks.\\

\begin{table}[t]
\small
\centering
\caption{Vulnerability coverage percentages across VMs.}
\begin{tabular}{lccccc}
\hline
\textbf{Machine} & \textbf{VM1} & \textbf{VM2} & \textbf{VM3} & \textbf{VM4} & \textbf{Average} \\
\hline
PentestGPT & 83.33 & 58.33 & 91.67 & 48.15 & 70.37 \\
AutoPentester & \textbf{100.0} & \textbf{100.0} & \textbf{100.0} & \textbf{92.59} & \textbf{98.14} \\
\hline
\end{tabular}

\label{tab:vuln_cov}
\end{table}

\begin{table}[t]
\footnotesize
\centering
\caption{Performance Comparison on VMs.  Here, P and AP relate to PentestGPT and AutoPentester, respectively.}
\begin{tabular}{llp{1cm}p{1cm}p{1.2cm}p{1.2cm}}
\toprule
\textbf{VM} & \textbf{Tool} & \textbf{Steps ($\downarrow$)} & \textbf{Loops ($\downarrow$)} & \textbf{Human Inter. ($\downarrow$)} & \textbf{Incomplete Cmd. ($\downarrow$)} \\
\hline
VM1 & P & 11.33 & 3.67 & 22.33 & 10.00 \\
    & AP & \textbf{5.33} & \textbf{0.00} & \textbf{0.00} & \textbf{0.00} \\
\midrule
VM2 & P & 12.00 & 4.33 & 15.67 & 3.67 \\
    & AP & \textbf{5.67} & \textbf{0.00} & \textbf{1.00} & \textbf{0.00} \\
\midrule
VM3 & P & 11.33 & 2.33 & 18.00 & 6.67 \\
    & AP & \textbf{6.67} & \textbf{0.00} & \textbf{0.00} & \textbf{0.00} \\
\midrule
VM4 & P & 12.00 & 2.67 & 16.67 & 4.67 \\
    & AP & \textbf{11.00} & \textbf{0.67} & \textbf{0.00} & \textbf{0.00} \\
\midrule
Avg & P & 11.67 & 3.25 & 18.17 & 6.25 \\
    & AP & \textbf{7.17} & \textbf{0.17} & \textbf{0.25} & \textbf{0.00} \\
\midrule
\end{tabular}
\label{tab:VM_comparison}
\end{table}

% \begin{figure}[ht]
%     \centering
%     \includegraphics[width=1\linewidth]{./Figures/vm_results.pdf}
%     \caption{Results comparison on VMs. Note that AutoPentester achieves a higher vulnerability coverage, as shown in Table~\ref{tab:vuln_cov} while having a lower number of steps, loops, and human interactions.}
%     \label{fig:vm}
% \end{figure}

% \begin{figure}[ht]
%     \centering
%     \includegraphics[width=0.9\linewidth]{./Figures/VM_vul_cov.pdf}
%     \caption{Vulnerability coverage comparison on VMs. Note that AutoPentester achieves a higher vulnerability coverage while having a lower number of steps, loops, and human interactions, as shown in Table~\ref{tab:VM_comparison}.}
%     \label{fig:vm_2}
% \end{figure}

\noindent{\textbf{Runtime Analysis:} We compare the time and costs of AutoPenstester and PentestGPT when solving the HTB machines as shown in the Table~\ref{tab:HTB_comp}. For experiments, we used a computer with Intel Core i9 (2.8GHz) processors and 32GB RAM without any GPUs. Here we measure the time taken for each machine in minutes and the cost in dollars by multiplying the token usage by the GPT4-turbo model charging rates. Since AutoPentester has additional steps such as RAG, Results Verifier, and ACI, it consumes more tokens compared to PentestGPT. However, due to its non-repetitive workflow (fewer steps), its cost was only \$3.86 higher on average than PentestGPT's across HTB machines, while achieving a 12.74\% higher subtask completion rate. Furthermore, when factoring in the significantly high cost of a professional Pentester required by PentestGPT (compared to token overhead), AutoPentester proves to be more cost-effective. Autopentester takes longer (on average, 71.9\%) to complete than PentestGPT. This is due to the generalizability issues in automation. For instance, tools like Metasploit operate via an interactive CLI, where the response times of different exploits vary. As a result, the ACI must incorporate extended waiting periods, even for quick commands, ultimately prolonging the overall process. However, its automated nature enables execution outside of regular working hours, thereby reducing the actual time required to complete tasks.}

\subsection{Qualitative Analysis}\label{qualiitative}

%Next, we present the results of our user study. 

To capture the actual impact of AutoPentester in the cybersecurity industry, we did a survey as described in Section~\ref{sec:P_metric}. Figure~\ref{fig:survey} presents a comparative analysis of AutoPentester and PentestGPT based on responses from cybersecurity professionals. The survey uses a scale where 5 represents ``Excellent" or ``Strongly Agree," while 0 denotes ``Extremely Negative" or ``Strongly Disagree." The graph contains the mean values obtained for the following nine questions after merging repetitive questions asked for both HTB and VM reports. \\

%industry security professional based on the reports generated by each tool. Here, scale 5 is aggrited to Excellent or Strongly agree conditions while scale 0 is aggrited to Extreamly negative or Strongly disagree conditions.

\noindent{{\bf Q1}  -  The amount of the attack surface covered is satisfactory.}
 
\noindent{{\bf Q2} - The amount of information revealed is on par with an actual pentesting process.

\noindent{{\bf Q3} - Quality/advancement of the steps are on par with actual pentesting steps.

\noindent{{\bf Q4} - When solving HTB machines, the strategy of the steps (i.e., the logical connection between steps) is satisfactory.

\noindent{{\bf Q5} - Efficiency of the tools (considering number of repetitive/unnecessary steps taken) is satisfactory?

\noindent{{\bf Q6} - How much do these steps align with a Human Pentester?

\noindent{{\bf Q7} - Quality of the remediations provided

\noindent{{\bf Q8} - Clarity of the information provided is satisfactory

\noindent{{\bf Q9} - If you use these tools for your professional work, will they save you time in basic penetration testing \\ 

\begin{figure}[t]
    \centering
    \includegraphics[width=1\linewidth]{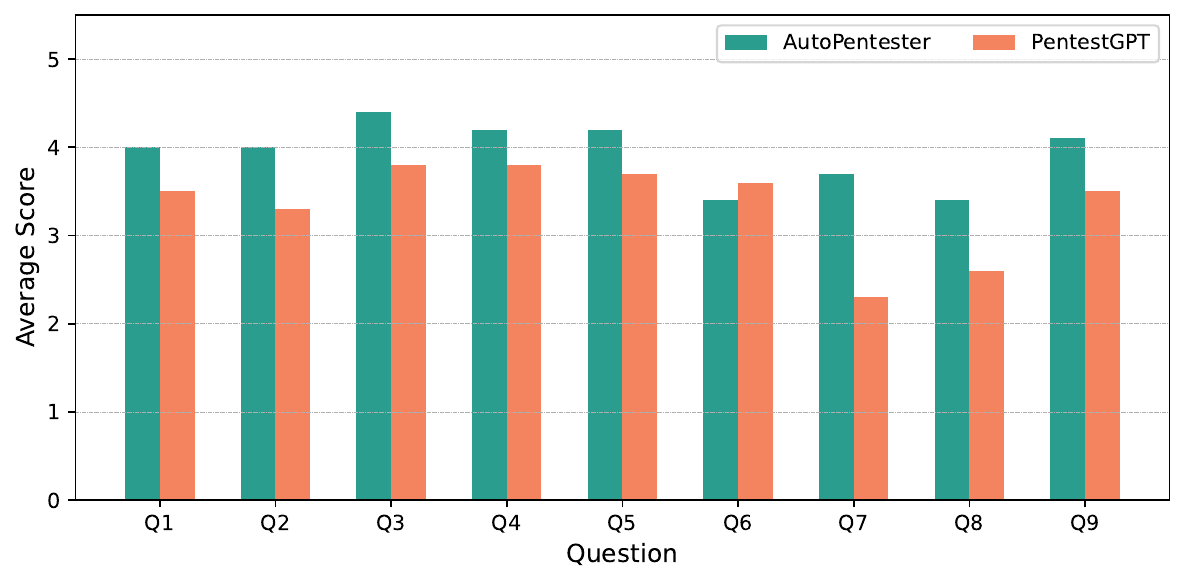}
    \caption{Results of the user study.}
    \label{fig:survey}
\end{figure}

According to the results for Q1 and Q2, AutoPentester covers a higher attack surface and gathers more relevant information compared to PentestGPT. Results for Q3 show that AutoPentester takes more advanced steps at the level of actual pentesting.

According to the results for Q4, the strategies used by AutoPentester when finding the attack vector in HTB machines are stronger than those of PentestGPT. Furthermore, AutoPentester is more efficient (Q5), provides good remediation steps (Q7), and clearer information (Q8) compared to PentestGPT. Furthermore, the professionals are more positive about AutoPentester saving time in professional red teaming tasks (Q9) than PentestGPT. Overall, AutoPentester has achieved a score of around 4, indicating a higher confidence compared to PentestGPT (mean 3.3) across all the questions.
%for all the questions, which indicates that it can be a valuable tool for the industry.
Finally, PentestGPT scored slightly higher than AutoPentester on Q6, as participants felt that, in certain situations, having flexibility in adjusting the strategy was preferable. \\ 

%PentestGPT has a slightly higher score for Q6 compared to AutoPentester. This is because participants thought that in some situations, participants prefer to have some flexibility in changing the strategy. 

%Furthermore, participants mention that due to the manual execution of suggested steps, it can be more flexible than the AutoPentester. The response "AutoPentester is not as flexible as human testers in adjusting strategies and responding to test results in real-time. In PentestGPT, humans can give input." states that. Consequently, in Q6, PentestGPT has a slightly higher score than the AutoPentester.

\begin{figure*}[ht]
    \centering
    \includegraphics[width=0.95\linewidth]{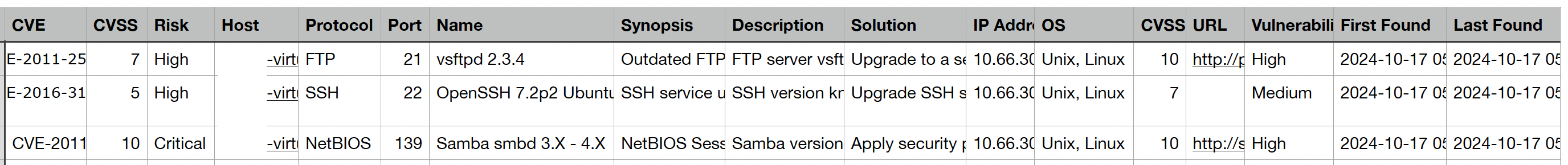}
    \caption{AutoPentester report: Summarized findings view.}
    \label{fig:csv}
\end{figure*}

\begin{figure*}[ht]
    \centering
    \includegraphics[width=0.9\linewidth]{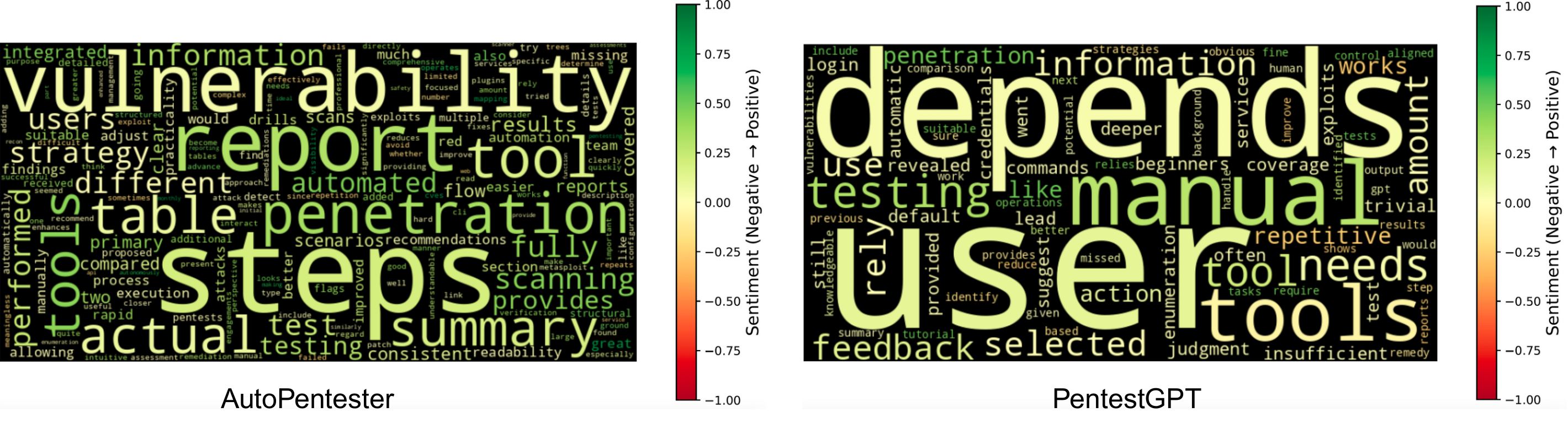}
    \caption{Word cloud: Free text analysis results. Here, font size rectifies the frequency of occurrence, and the color scale rectifies the sentiment context (positive or negative) of a particular word. Note that AutoPentester has more green faded words, indicating higher positive sentiment compared to PentestGPT.}
    \label{fig:word_cloud}
\end{figure*}

\noindent{\bf Free text analysis: } Survey participants provided valuable feedback, aiding both the comparison of AutoPentester and PentestGPT and the design of future solutions.

  AutoPentester was widely praised for its structured, logical workflow and clear remediation steps, making it ideal for enterprise assessments and red team drills. Its automated process was noted for significantly reducing manual effort in red team drills (\textit{``Fully automated process significantly reduces manual time, especially for rapid vulnerability scanning in large-scale red team drills''}). Additionally, participants valued the importance of conveying technical risks to non-technical stakeholders by incorporating executive summaries and impact assessments in reports, which is a feature in AutoPentester as shown in Figure~\ref{fig:csv}. On the other hand, participants observed that PentestGPT provides more manual flexibility but relies heavily on the user’s selection of commands and tools, sometimes leading to trivial or repetitive findings. 

To summarize the free-text analysis, we generated two separate word clouds for AutoPentester and PentestGPT, as illustrated in Figure~\ref{fig:word_cloud}. The comments were first carefully categorized based on the tool they referred to. Subsequently, the \textit{\texttt{TextBlob}} library~\cite{textblob_2018b} was used to perform contextual sentiment analysis and assign sentiment values to each word. Then we remove the stopwords and plot the remaining words, mapping their frequency to the font size. As visualized in the figure, AutoPentester contains a higher percentage of green-shaded words than PentestGPT, reflecting a generally more positive sentiment in the user feedback. Furthermore, the presence of terms such as \texttt{'automated'}, \texttt{'steps'}, \texttt{'summary'}, and \texttt{'report'} in user interactions with AutoPentester suggests that users recognize the importance of automation and the value of concise reporting. In contrast, the frequent occurrence of terms such as \texttt{'manual'}, \texttt{'user'}, and \texttt{'depend'} in PentestGPT cloud indicates reliance on manual processes and user intervention within its workflow. 

\subsection{Ablation Study}\label{sec:ablation}

% \begin{table}[t]
%     \centering
%     \footnotesize
%     \renewcommand{\arraystretch}{1.2}
%     \caption{Ablation study for HTB Bashed.}
%     \begin{tabular}{l l c c c c}
%         \toprule
%         Machine & Config & Compl. (\%) & Steps & Loops & Inc. Com. \\
%         \midrule
        
%         Bashed    & B     & 25.00  & \textbf{4.33}  & 0.31  & 0.38 \\
%               & B+R   & 33.33  & 11.67 & 0.49  & 0.21 \\
%               & B+L   & 25.00  & 5.00  & 0.13  & 0.73 \\
%               & B+V   & 25.00  & 7.00  & 0.38  & 0.05 \\
%               & B+R+L & 50.00  & 6.67  & 0.00  & 0.15 \\
%               & B+R+V & 41.67  & 9.00  & 0.30  & 0.04 \\
%               & B+L+V & 33.33  & 4.67  & 0.00  & 0.07 \\
%               & \textbf{B+R+L+V} & \textbf{100.0} & 5.67  & \textbf{0.00}  & \textbf{0.00} \\
%         \bottomrule
%     \end{tabular}
    
%     \label{tab:ablation}
% \end{table}

\begin{table}[t]
    \centering
    \footnotesize
    \caption{Ablation study for HTB Lame.}
    \renewcommand{\arraystretch}{1.2}
    \begin{tabular}{l l c c c c}
        \toprule
        Machine & Config & Compl. (\%) & Steps & Loops & Inc. Com. \\
        \midrule
        Lame     & B     & 33.33  & 9.33  & 0.52  & 0.42 \\
              & B$^*$    & 41.67  & 9.00  & 0.48  & 0.43 \\
              & B$^*$+R   & 50.00  & 8.67  & 0.50  & 0.18 \\
              & B$^*$+L   & 41.67  & 7.33  & 0.05  & 0.50 \\
              & B$^*$+V   & 33.33  & 5.33  & 0.38  & 0.13 \\
              & B$^*$+R+L & 91.67  & 7.67  & 0.30  & 0.12 \\
              & B$^*$+R+V & 58.33  & 12.00 & 0.56  & 0.06 \\
              & B$^*$+L+V & 58.33  & 6.33  & 0.00  & 0.05 \\
              & \textbf{B$^*$+R+L+V} & \textbf{100.0} & \textbf{4.33}  & \textbf{0.00}  & \textbf{0.00} \\
        \bottomrule
    \end{tabular}
    \label{tab:ablation}
\end{table}

We conducted an ablation study to evaluate the contribution of each module to the overall performance. Following a similar architecture to PentestGPT, we consider a framework that consists only of Summarizer, Analyzer, and Generator as our baseline (B). The B$^*$ represents the same architecture but with the reasoning based Strategy Analyzer (replacing Analyzer) discussed in Section~\ref{sec:analyzer_ag}. Then we add RAG (R), Repition Identifier (L), and Results Verifier (V) to that developed baseline (B$^*$) and report the performance. We divide the number of Loops and the Incomplete Commands by the respective step count and report them in the per-step scale to make them independent from the step count and emphasize the value of each module. 

Table~\ref{tab:ablation} presents the results for the HTB machine Lame. As can be seen, the reasoning based Strategy Analyzer (B$^*$) improves the subtask completion rate by 25.0\% compared to the baseline (B), by deriving correct strategies using the previous findings stored in the PTT. Adding the RAG module (B$^*$+R) further enhances subtask completion by providing the Generator with relevant guidance to produce precise commands for a selected task. Furthermore, it helps to generate complete commands with the correct target IP, port numbers, and file paths. The Repetition Identifier (B$^*$+L) significantly reduces the loops per step by 90.5\% on average, reducing it from 0.48 to 0.05. The Results Verifier verifies the results and adjusts commands, reducing the incomplete command ratio by 80.14\% on average, reducing it from 0.43 to 0.13. When the two modules are added to the baseline, RAG and the Repetition Identifier combination (B$^*$+R+L) have the highest performance compared to the other two combinations, indicating that those two are the most critical modules.

% We repeated the ablation study for two other machines (HTB Lame and Metasploitable II) and observed similar trends. We have included those results in Appendix~\ref{sec:ap_ablation}.

\subsection{Failure cases}

Finally, we discuss failure scenarios of AutoPentester. Out of the 10 HTB machines, AutoPenstester fully completes the Lame and Bashed machines, failing at a subtask on the other eight machines due to the reasons presented in Table~\ref{tab:failure_reasons}. It failed to identify the correct strategy on 4 machines. Despite LLMs having substantial knowledge of cybersecurity tools and vulnerabilities, they lack the ability to always find the correct strategies to navigate through the attack path. In contrast to the VMs, where we enumerate each service for vulnerabilities, HTB machines need advanced strategies to identify the attack vector by analyzing the previous findings. For example, in the Authority HTB machine, AutoPentester does not identify the decrypted pwm\_admin\_password as the key to log in to the web service running on port 8443. Furthermore, AutoPentester does not automatically search for additional information, resulting in 2 other failures due to the lack of knowledge on required exploits. For example, the Topology HTB machine contains a vulnerability based on an equation generator in LaTeX, which requires additional knowledge through web browsing. Overall, AutoPentester struggles with web applications, failing in 2 subtasks. Since it tries to rely on curl commands, it thereby loses information and becomes unfocused due to large response texts.

%and reported results in the Appendix (\ref{tab:ablation_appendix}).

% \begin{figure}[h!]
%     \centering
%     \includegraphics[width=1\linewidth]{./Figures/verifier.png}
%     \caption{An example of Results Verifier adjusting the incorrect commands}
%     \label{fig:verifier_exmp}
% \end{figure}

\section{Discussion and Concluding Remarks}\label{sec:discussion}

\begin{table}[t]
\centering
\small
\caption{Failure reasons analysis.}
\begin{tabular}{lc}
\toprule
\textbf{Reason} & \textbf{Count} \\
\midrule
Couldn't identify the correct strategy & 4 \\
Missing knowledge about the required exploit & 2 \\
Failed to navigate on a GUI & 2 \\
\midrule
\end{tabular}
\label{tab:failure_reasons}
\end{table}

We proposed AutoPentester, an LLM agent-based framework for automated pentesting. Our performance evaluation on HTB and custom VMs showed that AutoPentester outperforms the PentestGPT baseline in nearly all tasks, offering greater autonomy, higher efficiency, and fewer errors. Specifically, it achieves a 27.0\% higher subtask completion rate and 39.5\% greater vulnerability coverage while requiring 18.7\% fewer steps and 92.6\% less human intervention. Its CoT-based Strategy Analyzer improves the correct navigation through the attack vector while the PentestGPT baseline heavily relies on human feedback for identifying strategies, as evident in their demonstration video (frame 4:16 of~\cite{gelei_deng_2023_L}). The user survey highlighted the benefits of AutoPentester's structured approach to automation, making it suitable for enterprise security assessments and large-scale red team drills. Its ability to interact with the CLI allows testers to multitask; however, experts caution that automated exploitation requires careful handling in production environments. \\

\noindent{\textbf{Limitations}: In fully automated mode, AutoPentester relies on CLI tools like curl to interact with GUI-based interfaces (e.g., web apps), which makes task execution challenging. As a mitigation, we support interactive mode where the user can follow the steps proposed by the AutoPentester and give feedback (observations) as text. 
%does not perform well with GUI interfaces (such as web applications) in the fully automated mode since it is majorly based on CLI commands such as curl and GET/POST requests. In the current setting, for web applications, we follow the steps proposed by the AutoPentester in the interactive mode and give feedback (observations) as text. 
Furthermore, the commands generated by the generator tend to focus on the content suggested by the RAG, sometimes limiting its scope and missing corner cases, such as using a specific GitHub repository for an exploit. Consequently, maintaining an up-to-date knowledge base is critical to ensure higher performance. Additionally, the user study would have benefited more by a larger sample size. However, we highlight that it is extremely difficult to find industry professionals to volunteer for these types of studies.}\\

\noindent{\bf Future research directions:} As mentioned in Section~\ref{sec:discussion}, current LLMs cannot identify the complex strategies in penetration testing tasks. Therefore, fine-tuning LLMs to identify strategies in pentesting would largely benefit this research domain. Similar to recent advancements in LLM reasoning and strategy learning in games~\cite{mnih2013playing}, Reinforcement Learning (RL) and its variants, such as Reinforcement Learning from Human Feedback (RLHF)~\cite{bai2022training} and Direct Preference Optimization (DPO)~\cite{rafailov2023direct}, can be used to fine-tune LLMs to learn strategies in pentesting tasks. Furthermore, as cyber professionals suggest, the addition of a GUI interaction module and web-focused tools like ZAP and OpenVAS will elevate the capabilities of automated pentesting tools.

\section{Acknowledgement}

This research was supported by the Australian Government through the NSW Connectivity Innovation Network Fund.

%\section*{References}

\bibliographystyle{IEEEtran}
\bibliography{Sections/bibliography}

\appendices

\end{document}